\documentstyle [twoside,12pt]{article}
\begin{document}
\setcounter{page}{1}
\setlength{\baselineskip}{.7cm}
\sloppy
\begin{center}
{\bf TWO EXTENDED VERSIONS OF THE CONTINUOUS 2D-HEISENBERG MODEL}
\end{center}
\begin{center}
E. Alfinito, G. Profilo and G. Soliani
\end{center}
\begin{center}
Dipartimento di Fisica dell'Universit\`a and Sezione INFN, 73100 Lecce,
Italy.
\end{center}
\medskip
\medskip
\medskip
\medskip
\begin{abstract}
We analyze two extended versions [the Ishimori model (IM) and a related
system, which will be called modified Ishimori model (mIM)] of the
continuous Heisenberg model in (2+1)-dimensions within the complex Hirota
scheme.
The IM, proposed in 1984, is an integrable (2+1)-dimensional topological
spin field model which has been studied in many theoretical frameworks. The
mIM has been introduced quite recently by some of the present authors
[Phys. Rev. {\bf B 49}, 12915 (1994)].  Using the same stereographic
variable in the Hirota formulation, we build up some new exact
solutions both for the IM and the mIM in the compact and noncompact case.
For the IM new configurations are a class of static solutions related to a
special third Painlev\'e equation, time-dependent solutions linked to an
other kind of the third Painlev\'e transcendent, and asympotic time-dependent
solutions whose energy density behaves as Yukawa potential. For the mIM, new
configurations are a class of exact solutions expressed in terms of
elliptic functions, and a class of time-dependent solutions related to a
particular form of the double sine-Gordon and the double sinh-Gordon
equations with variable coefficients. We discuss the new
configurations and certain known solutions which clarify the different
possible phenomenological role
played by the considered topological spin field models.
\end{abstract}

\bigskip

{\bf I. INTRODUCTION}

\par Topological field models, usually relevant to string theory
${\displaystyle{\cite{WI}}}$,
have been applied successfully in the last years to handle many problems
pertinent to condensed matter physics ${\displaystyle{\cite{AC}}}$.
\par Here we consider the topological spin field model in two-space and
one-time
dimensions:

$$S_t = {1\over {2i}}[S, S_{xx} + \alpha^2S_{yy}] + \beta^2S_y\phi_{x} - S_x
\phi_y,
\eqno(1.1a)$$

$$\phi_{xx} + \alpha^2\beta^2\phi_{yy} = 2\alpha^2\beta^2Q,\eqno(1.1b)$$

where $\alpha^2 = \pm1$, $\beta^2 = \pm1$, subscripts denote partial
derivatives, $S=S(x,y,t)$ is a 2$\times 2$ matrix defined by

$$S=\left(\matrix{S_3&\kappa S^*_+\cr
          \kappa S_+&-S_3\cr}\right),\eqno(1.2)$$

$S_+ = S_1 + iS_2$, the asterisk means complex conjugation,
$Q={1\over 2}{\{Tr(i/2)S[S_x,S_y]\}}$ is a conserved topological charge
density, and $\phi$ is a real scalar field. The functions $S_j(x,y,t)$
(j=1,2,3), which
are real-valued components of a classical unit "spin" vector $\vec S(x,y,t)
$, belong to the two-dimensional sphere $S^2$ $(\kappa^2 = 1)$, or the
pseudosphere $S^{1,1}$ $(\kappa^2 = -1)$, i.e.

$$S_3^2 + \kappa^2(S_1^2 + S_2^2) = 1.\eqno(1.3)$$

\par For $\beta^2 = -1$, Eqs.(1.1) describe the Ishimori model (IM), which
can be regarded as an integrable version (it has a Lax
pair formulation ${\displaystyle{\cite{IS,KO}}}$) of the continuous
2D-Heisenberg model ${\displaystyle{\cite{MA}}}$.
Both the compact $(\kappa^2 = 1)$ and the noncompact $(\kappa^2 = -1)$ IM
admit exact solutions classified by an integer topological charge (
localized solitons ${\displaystyle{\cite{DU}}}$, vortex-like
${\displaystyle{\cite{IS}}}$ and closed string-like
configurations ${\displaystyle{\cite{LE}}}$). Furthermore, the IM possesses an
infinite dimensional
symmetry algebra of the Kac-Moody type with a loop algebra structure
${\displaystyle{\cite{PR,KA}}}$.
This feature characterizes other nonlinear field equations in 2 + 1
dimensions of physical significance having a Lax pair formulation, such as
the Kadomtsev-Petviashvili equation ${\displaystyle{\cite{DA}}}$,
the Davey-Stewartson equation ${\displaystyle{\cite{CH}}}$
and the Three-Wave Resonant system
${\displaystyle{\cite{RA}}}$. Apart from these nice
properties, at present it is not known whether the IM is a Hamiltonian
system.
\par Conversely, for $\beta^2 = +1$, Eqs.(1.1) describe a spin field system
endowed with a Hamiltonian structure
${\displaystyle{\cite{LM}}}$. We shall call this system a
modified Ishimori model (mIM). Similarly to what happens for the IM, the
mIM allows a symmetry algebra of the Kac-Moody type with a loop algebra
structure. However, this does not imply that the mIM is surely integrable.
In fact, so far a Lax pair has been found only for $\phi_{xy} = 0$
${\displaystyle{\cite{LM}}}$. The
question of the integrability of the mIM for $\phi_{xy}\ne0$ remains open.
\par Just as it occurs for the IM, the mIM provides similar solutions, a
few of them turn out to be of the helical and the roton-type, and meron-
like configurations provided by a fractional topological charge
${\displaystyle{\cite{LM}}}$.
These results indicate that the IM and
the mIM may refer to different physical situations. This appears mostly
evident in relation to the configurations endowed with a nonvanishing
topological charge. In fact, the vortices found in the compact IM
${\displaystyle{\cite{IS}}}$ and
the string-like configurations allowed by its noncompact version
${\displaystyle{\cite{LE}}}$,
have an {\it integer} topological charge, while, as we shall see later, the
meron-like excitations in mIM are characterized by a {\it fractional}
topological charge.
\par The above considerations suggest that it should be interesting to
pursue a comparative study of the IM and the mIM models. Keeping in mind
this idea, in the following we apply the Hirota representation to look
for a special class of
configurations by choosing the same form of the stereographic variable in
terms of which one can express the spin field components and the auxiliary
field $\phi$. In doing so, for the (compact and non-compact) IM we obtain
some new static and dynamical configuration which can be expressed in
terms of certain special forms of the third Painlev\'e transcendent. Other
interesting new time-dependent configurations lead to asymptotic
expressions for the spin-field variables which are associated with an
energy density of the Yukawa type.
\par On the other hand, for the non-compact mIM we find as new static
exact configuration, a class of solutions expressed in terms of elliptic
functions. (A correspondig class of configurations for the compact mIM has
been already determined in Ref.[13]). Furthermore for both the compact and
the non-compact mIM an other new result is constituted by a class of
time-dependent solutions related, respectively to a particular form of the
double sine-Gordon and the double sinh-Gordon equations with variable
coefficients.
\par For our purposes, let us recall the Hirota scheme
${\displaystyle{\cite{HI}}}$. This
consists essentially in writing Eqs.(1.1) by using the stereographic
projection representation

$$S_+ = {2\zeta \over {1 + \kappa^2|\zeta|^2}}, \quad S_3 = {{1 - \kappa^2|
\zeta|^2}\over {1 + \kappa^2|\zeta|^2}},\eqno(1.4)$$

and putting $\zeta ={g\over f}$, where $f = f(x,y,t)$ and $g = g(x,y,t)$
are two arbitrary differentiable complex functions. Then, Eqs.(1.1) take
the form

$$(|f|^2 - \kappa^2|g|^2)(iD_t - D_x^2 - \alpha^2 D_y)(f^* \cdot g) - f^* g
(iD_t - D_x^2 - \alpha^2 D_y^2)(f^* \cdot f - \kappa^2 g^* \cdot g) = 0,
\eqno(1.5a)$$

$$\phi_{xx} + \alpha^2 \beta^2 \phi_{yy} = {{4i\alpha^2 \beta^2 \kappa^2}
\over {\Delta^2}}[D_y(g \cdot f) D_x(g^* \cdot f^*) - D_y (g^* \cdot f^*)D_x
(g \cdot f)],\eqno(1.5b)$$

with $\Delta = |f|^2 + \kappa^2|g|^2$, where the operators $D_t$, $D_x$,
and $D_y$ stand for the "antisymmetric derivatives", i.e. $D_t (a \cdot b) =a_t
b - a b_t$, and so on.
\par A particular solution to Eq.(1.5b) valid for any value $(\pm1)$ of the
parameters $\alpha^2$ and $\beta^2$, is given by

$$\phi_x = {-2i\beta^2 \alpha^2\over \Delta}D_y(f^*\cdot f + \kappa^2 g^
* \cdot g), \quad \phi_y = {2i\over \Delta}D_x (f^* \cdot f + \kappa^2 g^
* \cdot g).\eqno(1.6)$$

However, the compatibility condition $\phi_{xy} = \phi_{yx}$ is not
identically satisfied. Therefore, this is a constraint which has to be taken
into account in order to solve Eq.(1.5a).
\par Interesting phenomenological aspects of the IM and the mIM can be
evidenced assuming first that $f$ and $g$ are (complex) functions of $z=x +iy$
and its conjugate, namely $f = f(z,z^*,t)$ and $g = g(z, z^*,t)$.
Consequently, with the help of the operators $\partial_z = {1\over
2}(\partial_x - i\partial_y)$ and $\partial_{z^*} = {1\over 2}(\partial_x + i
\partial_y)$, Eqs.(1.5), (1.6) and the related compatibility condition can
be written in complex form. We shall call the full set of these equations
complex Hirota's formulation $(CHF)$ of the spin field model (1.1) (see the
Appendix).
Second, we are looking for special solutions to the $CHF$ by setting
$f = [a^*(z^*)]^{1/2}$ and $g = [a(z)]^{1/2}\psi(|z|)$, where $a(z)$ and
$\psi(|z|)$ are, respectively, a complex and a real function to be determined.
This choice corresponds to the stereographic variable

$$\zeta = [a(z)/a^*(z^*)]^{1/2}\psi(|z|).\eqno(1.7)$$

\par To be precise, below we shall limit ourselves to the cases $\alpha^2
=1$, $\kappa^2 = \pm1$.\vskip 1cm

{\bf II. CASE $\beta^2=-1$ }\vskip 0.5cm
\par Let us put Eqs.(1.5), for $\beta^2 =-1$ (IM), in the CHF.
Then, the compatibility condition
$\phi_{zz^*} =\phi_{z^*z}$ (see (1.6)) entails:
$$a(z)=a_0 exp[(\lambda/2)z^2], \eqno(2.1)$$
where $a_0$ and $\lambda$ are, respectively, an arbitrary complex and a
real constant. On the other hand, the complex form of (1.5a) furnishes the
nonlinear ordinary differential equation:
$$(1+\kappa^2 \psi^2)(\psi_{rr} + {1\over r} \psi_r) + (1-\kappa^2 \psi^2)
|{a_z\over a}|^2 = 2
\kappa^2 \psi \psi_r^2, \eqno(2.2)$$
where $ a(z) $ is given by (2.1) and $z= r e^{i\theta}$. Using the
transformation
$$ u = ln r , \qquad i)\;  \psi = tan{\gamma\over 4},\quad for \quad\kappa^2
= 1; \quad ii)\;\psi=tanh{\gamma\over 4},\quad for \quad \kappa^2 = -1,
\eqno(2.3)$$
Eq. (2.2) takes, correspondingly, the form:

$$\gamma_{uu} + \lambda^2 e^{4u} sin\gamma=0 \eqno(2.4)$$
and
$$\gamma_{uu} + \lambda^2 e^{4u} sinh\gamma=0. \eqno(2.5)$$

Equations (2.4) and (2.5) are related to a special case of the third
Painlev\'e transcendent, defined by ${\displaystyle{\cite{DV}}}$ :
$$ {d^2W\over dz^2} = {1\over W} ({dW\over dz})^2 - {1\over z} {dW\over dz}
+ (\alpha_0 W^2 + \alpha_1)+ \alpha_2W^3 +{\alpha_3\over W}, \eqno(2.6) $$
where $ W=W(z)$, and $\alpha_j \quad (j=0,1,2,3)$ are arbitrary
constants.
\par This can be seen by putting in (2.4) and (2.5):
i) $ e^{2u} = \sigma$, $ W = e^{i{\gamma\over 2}}$, and ii)
$ e^{2u} = \sigma$, $W = e^{\gamma\over 2}$, respectively. We get
$$ W_{\sigma\sigma} = {W_\sigma^2 \over W} - {1\over \sigma} W_\sigma
- {\lambda^2\over 16} (W^3 -{1\over W}). \eqno(2.7)$$
Thus, Eq. (2.7) corresponds to the particular case of the third Painlev\'e
equation (2.6) where $\alpha_0 = \alpha_1 = 0 $ and $\alpha_2 = -\alpha_3
=-{\lambda^2\over 16}. $
\par Dynamical configurations to the IM yielding (2.7) when the time is
switched off can also be obtained. Indeed, starting from
$$ \zeta^\prime = \zeta \rho(t), \eqno(2.8)$$
where $ \zeta$ is given by (1.7) and $\rho(t)$ is a function of the time to
be  found, the CHF provides $\rho = exp[-i(Et+D)]$ with E, D real constants,
and
$$ (1+ \kappa^2 \psi^2)(\psi_{rr} + {1\over r} \psi_r) -2 \kappa^2 \psi
\psi_r^2 +
\lambda^2 r^2(1- \kappa^2 \psi^2)+E (1+ \kappa^2 \psi^2)\psi = 0.
\eqno(2.9)$$
By means of the substitution $\psi = tan{\gamma\over 4}$ (for $\kappa^2 = 1$),
 or $\psi = tanh{\gamma\over 4}$ (for $\kappa^2 = -1$),
Eq.(2.9) becomes
$$ \gamma_{uu} +\lambda^2 e^{4u}sin\gamma + 2E e^{2u}sin{\gamma\over 2} = 0
\quad (\kappa^2 =1), \eqno(2.10)$$
or
$$ \gamma_{uu} +\lambda^2 e^{4u}sinh\gamma + 2E e^{2u}sinh{\gamma\over 2} = 0
\quad (\kappa^2 =-1). \eqno(2.11)$$
\par The change of variables $ W = e^{i{\gamma\over 2}}$ and $e^{2u} =\sigma$
transforms Eq.(2.10) into the third Painlev\'e equation
$$ W_{\sigma\sigma} = {W_\sigma^2 \over W} - {1\over \sigma} W_\sigma
- {\lambda^2\over 16} (W^3 -{1\over W}) - {E\over 8\sigma} (W^2 -1),
 \eqno(2.12)$$
which corresponds to the choice: $\alpha_0 = -\alpha_1 = -{E\over 8} $
and $\alpha_2 = -\alpha_3=-{\lambda^2\over 16} $
of the free parameters $\alpha_j$ present in (2.6).

\par By rescaling the independent variable $\sigma$, namely by setting
$ \xi = i{\lambda \over 4} \sigma$, Eq. (2.12) takes the form
$$ W_{\xi\xi} = {W_\xi^2 \over W} - {1\over \xi} W_\xi +
W^3 -{1\over W} + {2\nu\over \xi} (W^2 -1),\eqno(2.13)$$
with $\nu = {iE \over 4\lambda}$.

\par On the other hand, by taking ${\widetilde  W } = e^{\gamma\over 2}$ with
$e^{2u} =\sigma$, Eq.(2.11) reduces formally to Eq.(2.12), where now
${\widetilde W}$ is a real function.
Assuming $ \tau = {\lambda \over 4} \sigma$, we are led to the equation
$$ {\widetilde W}_{\tau\tau} = { {\widetilde W}_\tau^2 \over  {\widetilde W}} -
{1\over \tau} {\widetilde W}_\tau - {\widetilde W}^3 +{1\over {\widetilde W}} -
{2 {\widetilde \nu}\over \tau}( {\widetilde W}^2 -1),\eqno(2.14)$$
with ${\widetilde \nu} = - {E \over 4\lambda}$.
\par Equations (2.13) and (2.14) are invariant under the transformations

$$W\rightarrow {1\over W}, \quad  {\widetilde W}\rightarrow
{1\over{\widetilde W}},\eqno(2.15)$$

respectively.
\par At this stage some comments are in order.
\par {\it i)} Equation (2.13) coincides formally with the equation (1.31) of
Ref. 16 for the scaling limit of the spin-spin correlation function of the
two-dimensional Ising model. To be precise, in Ref. 16 one-parameter
family of solutions $\eta(\tau;\nu,\mu)$ to the above mentioned Eq.(1.31)
was found by the request that these remain bounded as the independent
variable $\tau$ approaches infinity along  the positive real axis.
Furthermore, the authors of Ref. 16 built up the large and the small-
$\tau$ behavior of $\eta(\tau;\nu,\mu)$ under certain conditions for the
parameters $\mu$ and $\nu$. For example, as $\tau\rightarrow\infty$ one has
$$\eta \sim 1 - \mu\Gamma(\nu + {1\over 2})2^{-2\nu}\tau^{-\nu - {1\over2}}
e^{-2\tau},\eqno(2.16)$$
where $\Gamma$ denotes the gamma function. This expansion will be used later
to provide explicit asymptotic solutions to the Ishimori model.
\par {\it ii)} Equation (2.14) resembles Eq.(1.31) of Ref. 16, but it is
really different from the latter because the term ${\widetilde W}^3 - 1/
{\widetilde W}$ in (2.14) has opposite sign. At present, the role of
Eq.(2.14) in the context of spin field models seems unknown. Its possible
physical
meaning could be explored following a procedure similar to that exploited
in Refs. 16.
\par Now, by substituting

$$\zeta^{\prime} =e^{i(\lambda xy - Et + \delta)}\psi(r)\eqno(2.17)$$

into (1.4), where $\delta$ is a constant, we get the spin field components
$S_j$ in terms of $\psi$ (see
(2.8), (1.7) and (2.1)):

$$S_1 = 2 cos(\lambda xy - Et +\delta){\psi\over{1 + \kappa^2\psi^2}},\eqno
(2.18a)$$

$$S_2 = 2sin(\lambda xy - Et +\delta){\psi\over{1 + \kappa^2\psi^2}},\eqno(
2.18b)$$

$$S_3 = {{1 - \kappa^2\psi^2}\over {1 + \kappa^2\psi^2}}.\eqno(2.18c)$$

Limiting ourselves, for simplicity, to consider the compact case ($\kappa^2
= 1$), the quantities (2.18) become

$$S_1 = cos(\lambda xy - Et +\delta) sin\gamma,\quad S_2 = sin(\lambda xy - Et
+
\delta) sin\gamma,\quad S_3= cos\gamma,\eqno(2.19)$$

where $sin\gamma = {1\over{2i}}(W - W^*)$,\quad $cos\gamma = {1\over{2}}(W +
W^*)$, and $W = W(\xi)$ satisfies Eq.(2.13).
   The auxiliary field $\phi$ can be derived from (1.6) keeping in mind
that

$g=e^{-i(Et + D)}[a(z)]^{1/2}\psi(r)$ and $f=[a^*(z^*)]^{1/2}$, $a(z)$
being expressed by (2.1). We get

$$\phi_x = -2\lambda xS_3, \quad \phi_y = -2\lambda yS_3,\eqno(2.20)$$

which furnishes the topological charge density (see( 1.16))

$$Q = {1\over 2}(\phi_{yy} - \phi_{xx}) = \lambda(xS_{3x} - yS_{3y}).$$

The total topological charge,

$$Q_T = {1\over {4\pi}}{\int_{-\infty}^{+\infty}\!\int_{-\infty}^{+\infty} Q
dx\,dy}\eqno(2.21)$$

can be evaluated, in principle, from the properties of the third Painlev\'e
transcendent $W$ defined by Eq.(2.13) (see (2.19)).
\par Another interesting, more explicit example of solution to the IM
related to the third Painlev\'e equation arises from (2.11) by choosing
$\lambda = 0$ and $E<0$.
In fact, in this case Eq.(2.11) can be written as

$$W_{\rho \rho} ={1\over W}W^2_{\rho} - {1\over {\rho}}W_{\rho} + W^3 -
{1\over W}, \eqno(2.22)$$

where $e^{\gamma\over 4}= W$, \quad $e^u = r = 2\rho/(|E|)^{1\over 2}$, and

$$\zeta = e^{i(|E|t - D)}\psi(r).\eqno(2.23)$$

We remark that (2.22), where $W$ and $\rho$ are real quantities, is
exactly Eq.(1.3) (for $\nu=0$) studied in Ref. 16 . Therefore, we can
exploit (2.16) to provide an explicit asymptotic solution to the IM. In
doing so, from (1.4) and (2.23) we find

$$S_1 = cos(|E|t - D) sinh{\gamma\over 2},\eqno(2.24a)$$

$$S_2 = sin(|E|t - D) sinh{\gamma\over 2},\eqno(2.24b)$$

$$S_3 = cosh{\gamma\over 2},\eqno(2.25c)$$

where  $sinh{\gamma\over 2} = {1\over 2}(W^2 - W^{-2})$ and
$cosh{\gamma\over2}={1\over 2}(W^2 + W^{-2})$.

By resorting to (2.16) with $\nu=0$ and identifying $\tau$ with $\rho$, as
$\rho\rightarrow\infty$ we have

$$W(\rho;0,\mu) \sim 1 - \mu(\pi)^{1\over 2} e^{-2\rho}{\rho}^{-{1\over 2}}.
\eqno(2.25)$$

Then, the spin field components (1.4) become

$$S_1\sim -2\mu(\pi)^{1\over 2}cos(|E|t - D)e^{-2\rho}{\rho}^{-{1\over 2}},
\eqno(2.26a)$$

$$S_2\sim -2\mu(\pi)^{1\over 2}sin(|E|t - D)e^{-2\rho}{\rho}^{-{1\over 2}},
\eqno(2.26b)$$

$$S_3 \sim 1,\eqno(2.26c)$$

where the parameter $\mu$ is real.
The auxiliary field $\phi$ related to (2.23) turns out to be a constant.
This can be seen from (1.6) with $f=1$ and $g=\zeta$ (see (2.23)). The
topological charge density vanishes. On the other hand, the energy density
${\cal E}$ carried by the spin components (2.16) is

$${\cal E} = {1\over 2}{\sum_{j=1}^3 S^2_{jr}}={1\over 8}\gamma^2_r cosh\gamma
={1\over 4}|E| W^2_{\rho} (W^2 + W^{-6}),\eqno(2.27)$$

where $W$ obeys the special Painlev\'e equation of the third kind (2.14),
 $S_{jr}\,=\,\frac{\partial S_{j}}{\partial r}$, $W_{\rho}\,=\,
\frac{\partial W}{\partial \rho}$.
With the help of (2.25), we obtain the asymptotic value

$${\cal E} \sim 2|E|{\mu}^2(\pi) e^{-4\rho}{\rho}^{-1}.\eqno(2.28)$$

The expression (2.28) tells us that, for large values of $\rho$, the
energy density of the spin configuration (2.24) is of the Yukawa type.
\vskip 1cm
{\bf III. CASE $\beta^2=1$}
\par By using the stereographic variable (1.7) in the CHF of the spin
field model (1.1) for $\beta^2 =1 $ (mIM), from the compatibility
condition $\phi_{zz^*} = \phi_{z^*z}$ (see (1.6)) we obtain
$$ a(z) = a_{0} z^{\lambda}, \eqno(3.1)$$
where $a_{0}$ is an arbitrary complex constant, and $\lambda$ is a real
number. On the other hand, with the aid of (3.1) the complex version of
(1.5a) yields
$$ \bigl(1+\kappa^{2} \psi^{2}\bigr)\bigl(\psi_{rr} + {1\over r} \psi_{r}
\bigr) + {\lambda^2\over r^2} \bigl( 1- \kappa^2 \psi^2\bigr)\psi = 2\kappa^2
\psi\psi^{2}_{r}. \eqno(3.2)$$
By way of change of variables $ u = ln r$, and i) $\psi = tan {\gamma\over
4}$, for $\kappa^2 =1$; ii) $ \psi = tanh { \gamma \over 4},$ for $\kappa^2 =
-1$, Eq. (3.2) can be written as
$$ \gamma_{uu} + \lambda^2 sin\gamma =0,  \eqno(3.3)$$
and
$$\gamma _{uu} + \lambda^2 sinh\gamma =0, \eqno (3.4)$$
respectively. The first is the equation for the pendulum, which admits the
solution
$$ \gamma = 2 arsin[k sn(\lambda u, k)]  \eqno(3.5) $$
where sn($\cdot$) denotes the Jacobian elliptic function of modulus $k$
$(0 \le k\le 1)$.
The spin field components, the auxiliary field $\phi$, the topological
charge density and the energy associated with (3.5) and their limit cases
$ (k = 0$ and $k =1) $ have been already discussed in Ref. 13.
\par However, for the reader's convenience, below we report the main results.
Let us take $\lambda = 1$ for simplicity. Then, by taking $ g=
\bigl ({1-dn(u,k) \over z^*}\bigr)^{1/2}$ and $f= \bigl( {1+dn(u,k) \over
z}\bigr)^{1/2}$, where $dn^{2}(\cdot)-1 = k^2 sn^2(\cdot)$, the variable
(1.7) reads
$$ \zeta = \bigl( {z \over z^* }\bigr)^{1/2} \bigl[{ 1- dn(u,k) \over 1+
dn(u,k) }\bigr]^{1\over 2}. \eqno(3.6)$$
By introducing (3.6) into (1.4) gives the radially symmetric spin field
configuration
$$ S_1 = k sn(u,k)cos \theta,\quad S_2 = k sn(u,k)sin \theta,\quad S_3 =
dn(u,k),
\eqno(3.7)$$
while the auxiliary field $\phi$ turns out to be
$$ \phi = 2 arcsin[sn(u,k)] + \phi_0, \eqno(3.7')$$
$\phi_0$ being a constant of integration. The topological charge density
is
$$ Q = {1 \over r^2} {d\over du} dn(u,k), \eqno(3.8)$$
which implies a vanishing total topological charge $Q_T =0$.
Now, we recall that the mIM is a constrained Hamiltonian system described by
the Hamiltonian density ${\displaystyle{\cite{LM}}}$
$$ H = H_M + H_{\phi}= {1\over 2} \sum_{j=1}^3 \bigl( S_{jx}^2 + \alpha^2
S_{jy}^2 \bigr) +
{1\over 4} \bigl( \alpha^2 \phi_{x}^{2} + \phi_{y}^2 \bigr), \eqno(3.9)$$
where
$$H_M ={1\over 2} \sum_{j=1}^3 \bigl( S_{jx}^2 + \alpha^2
S_{jy}^2 \bigr),\qquad H_{\phi} = {1\over 4} \bigl( \alpha^2 \phi_{x}^{2}
+ \phi_{y}^2 \bigr), \eqno(3.9')$$
$$\phi_x = 2 \alpha^2 (q p_y -\Lambda_y),$$
$$\phi_y = -2(q p_x - \Lambda_x), \eqno(3.10)$$
$q$ and $ p$ are a pair of canonical variables defined by
$$q = - arctan{S_2\over S_1}, \quad p = S_3, \eqno(3.11)$$
and $\Lambda = \Lambda(x,y,t)$ is a differential function determined by
the compatibility condition $\phi_{xy}= \phi_{yx},$ namely
$$ \Lambda_{xx}+\alpha^2 \Lambda_{yy} = \partial_{x}(qp_x) + \alpha^2
\partial_y (qp_y). \eqno(3.12)$$
The quantities (3.10) obey Eq.(1.1b) ($\beta^2 =1$).
It is noteworthy that for $ \alpha ^2 =1$ (the case under consideration),
Eq.(3.12) takes the form
$$\nabla \cdot \vec v =0 , \eqno(3.13)$$
where
$$\vec v = \nabla\Lambda - q \nabla p. \eqno(3.14)$$
Therefore, the mIM ( for $\alpha^2 =1$) can be regarded as an
incompressible "spin fluid", in which the velocity is given by (3.14).
Formula (3.14) enables us to find the expression
for the velocity of the configuration (3.7), (3.7'). In doing so, from
Eq.(3.14), (3.10) and (3.11) we obtain
$$v_1 = {\phi_y \over 2} = {1\over r} sin \theta dn(u,k^2),$$
$$v_2 = -{\phi_x \over 2} =-{1\over r} cos \theta dn(u,k^2),  \eqno(3.15)$$
where $v_1$ and $v_2$ are the components of $ \vec v$ along the x and y-axis,
respectively.
\par We note that $|v|^2 \equiv H_{\phi}$ (see (3.9')). Then, the
contribution to the total energy density due to the field $\phi$ can be
interpreted essentially as the kinetic energy density of the configuration
(3.7), (3.7').
The nonlinear excitation (3.7) allows us to build up a configuration
endowed with a fractional topological charge. This can be done for $k=1$.
In fact, in this case $sn(u,k) \to tanh u $ and $ dn(u,k) \to sechu$.
Thus, from (3.7) we have
$$ S_1 = { r^2 - 1 \over r^2 + 1} cos \theta ,\quad  S_2 = { r^2 - 1 \over r^2
+ 1} sin \theta ,\quad S_3 = {2 r \over r^2 + 1} . \eqno(3.16) $$
\par The auxiliary field $\phi$ corresponding to (3.6) can be derived from
(1.6).
It reads ${\displaystyle{\cite{LM}}}$
$$\phi = 4arctan r ,  \eqno(3.17)$$
apart from a constant of integration.
\par The Hamiltonian density (3.9) related to the configuration (3.16) or
(3.17) becomes
$$ H = {1\over 2r^2} + {4 \over (1+r^2)^2}, \eqno(3.18)$$
where the terms at the r.h.s. are the contribution of the magnetic part and
of the field $\phi$, respectively.
\par The topological charge density is
$$ Q = {1\over 2}(\phi_{xx} + \phi_{yy}) = 2{1-r^2 \over r(1+ r^2)^2}.
\eqno(3.19)$$
\par Therefore, the total topological charge
$$Q_T ={1\over 4\pi}\int_{-\infty}^{+\infty}\int_{-\infty}^{+\infty}Q dxdy
\eqno(3.20)$$
vanishes.
Anyway, starting from (3.16), we can construct a static solution endowed
with a fractional topological charge $ Q_T = +{1\over 2}$, i.e.
$$ {\vec T} = \sigma (1- r) {\vec S} + \sigma (r- 1) {\vec S}_0,
\eqno(3.21)$$
where $\sigma$ stands for the step function, ${\vec S}$ is given by (3.16),
and ${\vec S}_0 \equiv (0,0,1)$.
\par A configuration having an opposite topological charge, $Q_T = - {1
\over 2}$, can also be found. This reads
$${\vec T^\prime} = \sigma (r- 1) {\vec S} + \sigma (1- r) {\vec S}_0.
\eqno(3.22)$$
The static solutions (3.21) and (3.22) bear some analogies with other
field configurations provided by a fractional topological charge
$(Q_T = \pm {1 \over 2})$, such as, for instance, the merons discovered in
the two-dimensional O(3) nonlinear $\sigma$-model and in the four-dimensional
non-Abelian gauge theory ${\displaystyle{\cite{GR}}}$.
\par Now let us deal with Eq.(3.4). This can be considered as an equation
of the sinh-Gordon (sinh-Poisson) type in the variable $u=lnr$. It is
related to the description of negative-temperature configurations in the
theory of vortex filaments in $^4He$. It affords the solution

$$\gamma = 4 arctanh[\sqrt{k} sn(v,k^{2})]. \eqno(3.23)$$

where

$$ v = \sqrt{{c-\lambda^{2}\over 8k}}(u-u_0), \eqno(3.24)$$

$c$ and $u_0$ are constants of integration, $ c >\lambda^2$, and $k$ is a
positive number such that
$$k = {c + 3\lambda^2 -\sqrt{8\lambda^2(c+\lambda^2)} \over c-\lambda^2}
< 1. \eqno(3.25)$$
The condition $c> \lambda^2$ ensures the reality of (3.23).
\par The spin field components can be obtained with the aid of (3.23) by
replacing into (1.4) the stereographic variable
$$\zeta = e^{i(\theta + \theta_0)\lambda}\;\sqrt{k}\; sn(v,k^2). \eqno(3.26)$$
(see (1.7), where $a(z)$ is given by (3.1) and $a_0 = |a_0| e^{i\theta_0}$).
Putting for simplicity $\theta_0=0$, these are
$$ S_1 = {2\sqrt{k}\; sn(v,k^2)cos(\lambda \theta) \over
1 - k\; sn^{2}(v,k^2)}, \eqno(3.27a)$$
$$ S_2 = {2\sqrt{k}\; sn(v,k^2)sin(\lambda \theta) \over
1 - k\; sn^{2}(v,k^2)}, \eqno(3.27b)$$
$$ S_3 = {1 +k sn^{2}(v,k^2)\over 1 - k sn^{2}(v,k^2)}. \eqno(3.27c)$$

on the other hand, by choosing $g(z)=\sqrt{a(z)}\;\psi(r)$ and $f(z^{*})=
 \sqrt{a^{*}(z^{*})}$, Eq.(1.6) yields

$$ \phi_{r} ={2\lambda \over r}\biggl( {1 +k sn^{2}(v,k^2)
\over 1 - k sn^{2}(v,k^2)}\biggr). \eqno(3.28)$$
and $\phi_{\theta} =0$.

By integrating (3.28), we get
$$ \phi(r) = 2\lambda \sqrt{8k \over c- \lambda^2}\;[2\Pi(k,v;k^2)- F(v;k^2)]
+ \;\;const, \eqno(3.29)$$
where $F(v;k^2)$ and $\Pi(k,v;k^2)$ denote the elliptic integral of the
first and the third kind, respectively, i.e.
$$ F(v;k^2)= \int^{sn(v,k^2)}_{0} {dt \over \sqrt{(1-t^2)(1-k^{2}t^{2})}},
\eqno(3.30)$$
and
$$\Pi(k,v;k^2)=\int^{sn(v,k^2)}_{0} {dt \over (1-kt^{2})
\sqrt{(1-t^2)(1-k^{2}t^{2})}}.\eqno(3.31)$$

\par Looking at (3.28) and (1.1b) the topological charge density is given
by
$$ Q= \phi_{rr} + {1\over r} \phi_{r}= {\lambda \over r} {d\over dr}
\biggl({ 1+k sn^{2}(v,k^2) \over  1-k sn^{2}(v,k^2)}
\biggr), \eqno(3.32)$$
which leads to a vanishing total topological charge.
With the aid of Eqs.(3.27) and (3.9), we can easily evaluate the energy of
the spin field configuration (3.28), (3.29). We shall omit here its
explicit expression.
\par Concerning the limit cases $k=0$ and $k =1$, only the
latter will be considered, because the former leads to a complex value of
$\gamma$.
The case  $k =1$, which corresponds to $\lambda =0$, yields (see (3.27))
$$ S_1 = cos\theta_{0} sinh[2(au+b)], \eqno(3.33a)$$
$$ S_2 = sin\theta_{0} sinh[2(au+b)], \eqno(3.33b)$$
$$ S_3 = cosh[2(au+b)], \eqno(3.33c)$$
where $a,\;b, \;\theta_0$ are real constants and $u = lnr$.
The auxiliary field $\phi$ turns out to be a constant and the topological
charge density $Q$ is zero (see (3.28) and (3.32)), while for the energy
density we obtain
$$ H = 2a^{2} e^{2u}\;cosh[4(au+b)]. \eqno(3.34)$$

\par In analogy to the IM, using Eq.(2.8), we can introduce special dynamical
solutions for the mIM as well. In this case the CHF furnishes
$$ \rho(t)= e^{-i(Et+D)},  \eqno(3.35a)$$
and
$$ \psi_{rr} + {\psi_{r} \over r} +{\lambda^2 \over r^2} \psi \biggl(
{1-\kappa^{2} \psi^2 \over 1+\kappa^{2} \psi^2}\biggr) +E\psi =
2 {\kappa^{2} \psi^{2}_{r} \psi \over 1+\kappa^{2} \psi^2}. \eqno(3.35b)$$
By means of the transformations $\psi = tan{\gamma\over 2}$ (for $\kappa^2 =
1)$
or $\psi = tanh{\gamma\over 2}$ (for $\kappa^2 = -1)$ and $u=lnr$, we get
$$ \gamma_{uu} + {\lambda^2 \over2} sin2\gamma + e^{2u} E sin\gamma =0
\qquad\qquad (\kappa^2 = 1),\eqno(3.36)$$
$$ \gamma_{uu} + {\lambda^2 \over2} sinh2\gamma + e^{2u} E sinh\gamma =0
\qquad\qquad (\kappa^2 = -1).\eqno(3.37)$$
These equations look as, respectively, a double sine-Gordon
and a double sinh-Gordon equation with variable coefficients.
Indeed, they resemble formally those obtained from (2.10) and (2.11)
by setting $\lambda =0$.
Therefore, in this case both the IM and the mIM allow configurations
having the same characteristics.
Finally, we notice that for small $\gamma$ Eqs.(3.36)
and (3.37) can be linearized to give equations of the Bessel type
${\displaystyle{\cite{CL}}}$.
\vskip 1cm
{\bf IV. CONCLUSIONS}
\vskip 0.5cm
We have investigated two extended versions of the continuous Heisenberg model
in 2+
1 dimensions using the Hirota technique. The first system is the Ishimori
model, while the second one has been introduced in
${\displaystyle{\cite{LM}}}$ and can be regarded
as a modified version of the former. The basic motivations for a
comparative study of these models are: {\it i)} the IM allows a Lax
pair representation, but it seems to be not endowed with an Hamiltonian
structure; {\it ii)} the mIM admits a Lax pair only for special values of
the auxiliary field. Conversely, it can be described by a Hamiltonian;
{\it iii)} the models can be formulated in an unified manner.
To the aim of clarifying the possible
phenomenological aspect of the systems, we have looked
for a class of solutions starting from the same {\sl ansatz} for the
stereographic variable $\zeta$ involved in the Hirota representation.
We have found new exact configurations for the two models under
consideration both in the compact and in the noncompact case. For the IM, these
are static and time-dependent solutions connected with certain particular
forms of the third Painlev\'e transcendent and a class of time-dependent
solutions whose asympotic behavior allows an energy density of the Yukawa
type. For the mIM, we have obtained a class of static solutions expressed
in terms of elliptic functions and time-dependent configurations related to
a particular form of the double sine-Gordon and the double sinh-Gordon
equations with variable coefficients.

On the basis of these results, it turns out that the IM and the mIM may
describe quite different physical
situations. This emerges in part from the comparison of Eqs. (2.4), (2.5),
(2.10) and (2.11) with Eqs. (3.3), (3.4) and (3.36), (3.37).
Furthermore, while the IM possesses vortex configurations labeled by an integer
topological charge ${\displaystyle{\cite{IS,LE}}}$, the mIM has meron-like
solutions which can be
interpreted as vortices ${\displaystyle{\cite{AF}}}$ characterized by a
fractional topological
charge.We remark also that for the mIM, which can be regarded as a
constrained Hamiltonian system, via (3.14) one can determine explicitly (
for $\alpha^2 = 1$) the velocity of the allowed excitations.
\par Finally, we notice that recently the continuous 2D-Heisenberg model
has been analyzed within the anyon theory ${\displaystyle{\cite{PA}}}$. It has
been shown that
static magnetic vortices correspond to the self-dual Chern-Simons solitons
described by the Liouville equation. The related magnetic topological
charge is associated with the electric charge of anyons. This result is a
challenge for scrutinizing, in this direction, both the Ishimori and its
modified version.

\vfill\eject

{\bf APPENDIX}
\vskip 1cm
\par Using the operators $\partial_z = {1\over 2}(\partial_x - i\partial_y)$
and
 $\partial_{z^{*}}={1\over 2}(\partial_x + i\partial_y)$, Eq. (1.5a) and the
compatibility condition $\phi_{xy} = \phi_{yx} $ (see (1.6)) can be written
, respectively as
$$ (|f|^2-  \kappa^2 |g|^2)\{i(f^{*}_t g - f^* g_t )-2(1+\alpha^2)
(g f^*_{zz^*}+ f^*g_{zz^*} -f^*_z g_{z^*}-f^*_{z^*} g_z) -$$
$$(1-\alpha^2)[g( f^*_{zz}+ f^*_{z^*z^*}) + f^*(g_{zz}+g_{z^*z^*})
-2f^*_{z}g_{z}-2f^*_{z^*}g_{z^*}]\} $$
$$- f^*g\{i[ff^*_t - f^*f_t -  \kappa^2(gg^*_t - g^*g_t)]- 2(1+\alpha^2)
[ff^*_{zz^*} + f^*f_{zz^*}-f^*_{z}f_{z^*} -f^*_{z^*}f_z $$
$$ - \kappa^2(gg^*_{zz^*} + g^*g_{zz^*} -g^*_{z}g_{z^*} - g^*_{z^*}g_z)]
-  (1-\alpha^2) [f(f^*_{zz} +f^*_{z^*z^*})-2f^*_{z}f_z -2f^*_{z^*}f_{z^*}
 + f^*(f_{zz} +f_{z^*z^*}) $$
$$-  \kappa^2(gg^*_{zz} +gg^*_{z^*z^*}-2g^*_{z}g_z -2g^*_{z^*}g_{z^*} +
g^*g_{zz} +g^*g_{z^*z^*})]\}=0, \eqno(A.1)$$
and
$$ 2(\alpha^2 + \beta^2) \{\Delta (f f^*_{zz^*}
+  \kappa^2 g g^*_{zz^*} - c.c.) - [(ff^*_z + \kappa^2 gg^*_z)
(ff^*_{z^*} + \kappa^2 gg^*_{z^*}) - c.c.]\}$$
$$ -(\alpha^2 - \beta^2)\{[f(f^*_{zz} + f^*_{z^*z^*} )
+ \kappa^2 g(g^*_{zz} + g^*_{z^*z^*}) - c.c. ]\Delta $$
$$+[(f^* f_z +  \kappa^2 g^* g_z)^2 + ( f^* f_{z^*} +
 \kappa^2 g^* g_{z^*})^2 -c.c. ] \} =0, \eqno(A.2) $$
with
$$ \Delta = (|f|^2 + \kappa^2 |g|^2).$$

\vfill\eject

\end{document}